# Learning Style Identification Using Semi-Supervised Self-Taught Labeling

Hani Y. Ayyoub and Omar S. Al-Kadi, Senior *Member*, IEEE

*Abstract*—Education is a dynamic field that must be adaptable to sudden changes and disruptions caused by events like pandemics, war, and natural disasters related to climate change. When these events occur, traditional classrooms with traditional or blended delivery can shift to fully online learning, which requires an efficient learning environment that meets students' needs. While learning management systems support teachers' productivity and creativity, they typically provide the same content to all learners in a course, ignoring their unique learning styles. To address this issue, we propose a semi-supervised machine learning approach that detects students' learning styles using a data mining technique. We use the commonly used Felder Silverman learning style model and demonstrate that our semi-supervised method can produce reliable classification models with few labeled data. We evaluate our approach on two different courses and achieve an accuracy of 88.83% and 77.35%, respectively. Our work shows that educational data mining and semi-supervised machine learning techniques can identify different learning styles and create a personalized learning environment.

*Index Terms*—Learning Management System, Personalized learning, Learning Style Model, Self-taught-labeling, Semi-supervised classification

## I. INTRODUCTION

Natural and human-caused disasters have had a significant impact on the education system, affecting students, educators, and organizations worldwide. Disruptions caused by events such as pandemics, wars, earthquakes, volcanoes, hurricanes, floods, and fires have forced a shift from traditional classroom instruction to distance learning, delivered through technology or online platforms. Cloud-based services and applications have enabled new opportunities for computer-based teaching and learning, with Learning Management Systems (LMS) being a prominent example of web-based educational frameworks. LMS provides a variety of useful online tools for assessment, communication, content uploading, and other interactive features, organizing e-learning content in one location with unlimited access. It also facilitates easy tracking of learners' progress, reducing learning and development costs and time. However, a limitation of LMS is its failure to consider different learning styles, treating all students equally.

The concept of learning styles has been a topic of interest in education for many years. It refers to the diverse ways in which individuals prefer to receive and process information [1]. By understanding a student's learning style, educators can better tailor their teaching methods to meet the needs of each individual student. The traditional way of detecting learning styles is through questionnaires that students fill out. While these questionnaires are considered reliable and valid [2], there is a risk of arbitrary answers that could lead to inaccurate results [3]. In addition, questionnaires are limited in their flexibility, which can decrease their validity. Participants may leave sections blank or answer questions in an inaccurate way. Furthermore, once a user model is initialized, it is difficult to revise any misclassification of a student's learning style. As a result, there is a need for more flexible and accurate methods to detect learning styles that can adapt to changes in a student's learning style over time.

In recent years, the integration of machine intelligence techniques into educational systems has garnered increasing interest [4]. E-learning systems generate vast amounts of data that can be collected and organized to gain valuable insights into student behavior and learning patterns. However, this approach may not capture all relevant aspects of information that can improve the quality of student outcomes. Hence, exploring other unseen relationships can offer significant benefits to curriculum management and teaching strategies. Automating the identification of students' learning styles using data mining and machine learning techniques is critical to developing an LMS that can adapt to different learning styles. Unlike the traditional approach of using questionnaires, which is static and provides limited flexibility, a machine learning approach can capture and analyze substantial amounts of data in real-time. Additionally, this approach has the potential to provide more accurate and reliable results, as it is less influenced by arbitrary student responses and can adapt to changes in a student's learning style over time. This personalized learning environment could improve the quality of educational services by providing a customized approach that adapts to students' different learning styles. By integrating machine intelligence techniques into educational systems, educational institutions can leverage data-driven insights to improve teaching strategies and enhance student outcomes.

In this work, self-taught machine learning technique is utilized to detect students' learning styles and provide a personalized learning environment on Moodle LMS [5]. However, this approach poses several challenges, particularly its compatibility with instructional design strategies. As designing a course is heavily influenced by the organization of its contents, activities, and context, it is important to consider different instructional design models such as ADDIE, Dick, and Carey model, etc., to ensure that the machine learning model fits well with the course structure [6]. Additionally, missing data such as forum activity can impact the quality of the model. To address these challenges, we developed a machine learning technique that employs a semi-supervised (i.e., self-training) method, which produces a reliable classification model even with small amounts of labeled data. By doing so, we aim to overcome the limitations of previous approaches and provide adaptive and personalized learning that meets the needs of various students. The paper aims to bridge the gap between personalized learning and instructional design strategies. Specifically, our work proposes the following contributions:

1) Assess student behavior from collected Moodle data to reliably label and infer students' learning style,
2) Employ a semi-supervised (self-training) mechanism to produce a reliable classification model from few labeled data,
3) Compare the performance of the semi-supervised training model using different classifiers to evaluate prediction ability on independent data,
4) Experiment with real data related to two different courses with students from diverse backgrounds and benchmark against similar work.

This paper is organized as follows: Section II serves as a background to learning styles and their significance in education. Section III provides an overview of relevant learning styles and discusses various approaches to detecting them. Section IV outlines our methodology, which consists of multiple phases, each described in detail. Section V, we present the experiments conducted using our proposed approach applied through Moodle LMS. Finally, Section VI summarizes our findings and provides directions for future research.

## II. BACKGROUND

In this section, we will provide a succinct overview of E-Learning and personalization, learning styles, and the ways in which LMS can be improved to facilitate personalized learning.

### A. E-Learning and Personalization

Learning in an educational environment encompasses two broad categories: traditional classroom learning and e-learning environment. Both environments share three fundamental components: a teacher, a student, and educational content. E-learning is an innovative approach to education that delivers electronically mediated, well-designed, student-oriented, and interactive learning experiences. These experiences are accessible from any location, at any time, using the internet and digital technologies that adhere to instructional design principles [7]. E-learning systems are web-based platforms, such as Content Management Systems and LMS. LMS is one of the most widely used e-learning systems, offering online tools for assessment, communication, content uploading, and other useful features. Additionally, LMS allows for content to be organized in one location, providing unlimited access to learners, while enabling teachers to track their progress with ease. This reduces the time and cost of learning and development.

Moodle [5], Blackboard [8], and Canvas [9] are some of the most used LMSs. However, LMSs have limitations in treating all students equally without considering their different learning styles [7]. To address this, LMSs can be customized with personalization features that provide a variety of learning objects to accommodate different learning styles. Personalization can be presented through the selection of learning content as well as customization of learning activities [10].

### B. Learning Styles

There are many learning style models available, with some of the most well-known being the Felder-Silverman learning style model (FSLSM) [11], Honey & Mumford [12], and Kolb's learning style model [13]. Each model offers distinct descriptions and classifications of learning styles.

Among these models, the FSLSM is particularly applicable to e-learning systems and is widely used. This model is based on the premise that students have preferences in terms of the way they receive and process information. It consists of four dimensions, each with two scales, as shown in Fig. 1.

The first dimension of the FSLSM is Sensing-Intuitive (S-I), which refers to the way students perceive information. The second dimension is Visual-Verbal (V-V), which relates to the way students prefer to receive information. The third dimension is Active-Reflective (A-R), which indicates how students prefer to process information. The fourth dimension is Sequential-Global (S-G), which refers to the way students organize information.

By understanding students' learning preferences through models such as FSLSM, e-learning systems can be personalized to provide content and activities that suit each student's individual learning style. This can lead to more effective and engaging learning experiences for students, resulting in improved learning outcomes.

### C. Adaptive learning

Adaptive learning systems can modify their behaviors and provide personalized learning objects for each student based on their individual characteristics, such as goals, knowledge, experience, background, and interests. These characteristics are referred to as the user model and vary from one student to another.

Several online adaptive systems have been developed that focus on the learner's learning [3, 14, 15]. To build a user model, there are two main techniques that can be used. The first is the *explicit model*, which is based on a questionnaire designed to assess different dimensions of the student's learning style. Once the learning styles have been identified, the system can provide appropriate learning objects that match the student's preferences. The second technique is the *implicit model*, which is based on observing the student's behaviors to identify their learning styles. This approach involves analyzing the student's interaction with the learning system, including their response time, frequency of use, and the types of learning objects they access. This information can then be used to create a user model that captures the student's learning preferences.

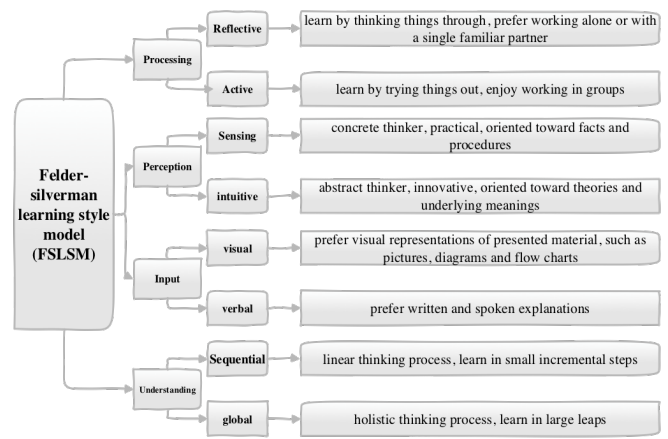

**Fig. 1:** The Felder-Silverman Learning Style Model.

Both techniques have advantages and disadvantages. The explicit model is more accurate, as it is based on self-reported data from the student. However, it can be time-consuming to administer and may not capture all aspects of the student's learning style. The implicit model is less intrusive, as it does not require the student to complete a questionnaire but may be less accurate if the system is unable to accurately interpret the student's behaviors.

*D. Personalized LMS*

Numerous studies have proposed enhancements to LMSs that support personalization and learner models based on individual learning styles. One such proposition involved the use of ontology in e-learning systems to model objects and relations based on the Felder-Silverman learning style model [16]. Imran et al. proposed a system that provides learners with personalized recommendations based on their navigational patterns and the learning objects visited by other learners with similar profiles, thereby enhancing the quality of learning [17].

Several studies have also been conducted to detect students' learning styles based on their behavior in LMSs using various techniques and approaches. In [17], students' learning styles was classified based on their navigational patterns and behaviors in an LMS, while another approach tested a Data Structure course using Moodle and collected data from the Moodle database about learners' behavior to detect learning styles depending on FSLSM [18]. Another study proposed an automatic approach to detect learning styles using the Index of Learning Styles (ILS) questionnaire and an analysis of learners' behavior when using Moodle [3]. However, there are technical-related issues in tracking user actions in Moodle, such as the inability to calculate time spent on each section or provide video statistics, which limits its ability to support personalized learning. To address this, we propose enhancements to Moodle that support personalization based on learning styles.

*E. Semi-Supervised learning*

Semi-supervised learning techniques act as a bridge between fully supervised and unsupervised learning, making use of both labeled and unlabeled data to boost model performance. In the context of detecting learning styles, there are several useful semi-supervised approaches. One such approach is self-training, where a model is trained on labeled data and then used to predict labels for unlabeled data. These predicted labels are then incorporated into the training process [19]. Another technique is co-training, which involves training multiple classifiers using different views of the data. These classifiers then label the unlabeled data examples, benefiting from each other's perspectives to improve predictions [20]. Graph-based methods construct graphs based on data similarities, with nodes representing instances and edges showing relationships. These methods propagate labels through the graph to predict labels for unlabeled data [21, 22]. Entropy minimization is also employed, aiming to reduce uncertainty in label predictions for unlabeled data by promoting more confident predictions and reducing the uncertainty in the model's predictions [23].

### III. LITERATURE REVIEW

In this section, we provide a review of previous work related to the research topic. Numerous studies have been conducted to detect students' learning styles using different techniques and approaches in LMS. Elfa Amir et al. identified two methods for detecting learning styles: collaborative approaches that use questionnaires and automated approaches that use learner behavior and actions in LMS [24].

El Fazazi et al. proposed an approach to automatically identify the learner's learning style based on web log files that contain learning behavior and then mapped to FSLSM [25]. They applied a fuzzy C mean algorithm and artificial neural network to predict the learning style, with a classifier accuracy of $79.1\%$. Abdullah et al. proposed a dynamic classification approach that identifies students' learning style according to FSLSM by extracting behavior and data from Moodle logs [18]. The classifier accuracy of this approach was $76\%$. In a similar work, Karagiannis and Satratzemi proposed an automatic approach to detect learning style to support adaptive courses in Moodle [3]. This approach uses the ILS questionnaire and analyzes learner behavior and actions in Moodle. The classifier accuracy for this approach was $70\%, 66\%, 75\%$, and $80\%$ for the four dimensions of FSLSM. Ferreira et al. proposed an approach to detect learning style and compared different machine learning algorithms [15]. They found that using a single classifier is not sufficient for all dimensions, and two classifiers combined provide a more accurate result.

García et al.[26] conducted a study to evaluate the efficacy of a Bayesian network model in discerning students' learning styles. The study aimed to employ Bayesian networks to pinpoint and grasp students' preferred learning modalities, thereby facilitating the customization of educational experiences. The findings held promise, as the model aptly identified students' perception styles; however, discrepancies surfaced when characterizing the understanding and processing dimensions of learning styles. Notably, the classification accuracy for this approach spanned across various dimensions of the Felder-Silverman Learning Style Model (FSLSM), achieving rates of $77\%, 63\%, 58\%$, and $52\%$.

Maaliw proposed an approach to detect learning style based on student behavior in a Moodle course, comparing Bayes network and decision tree classifiers [27]. The decision tree classifier (namely J480) had the highest average value of $89.91\%$. Liyanage et al. proposed an automatic prediction of learning style approach to detect learning style based on student behavior in Moodle courses [28]. They compared J48, Bayesian network, naïve Bayes, and random forests, with classifier accuracies of $70\%, 84\%, 91\%$, and $82\%$ for the four dimensions of FSLSM.

Hmedna et al. proposed predicting learners' learning styles based on their learning traces for the edX course "statistical learning" and used four machine learning algorithms [29]. The decision tree algorithm had the highest accuracy of over 98% value for three dimensions (i.e., Processing, Input, Understanding). Aissaoui et al. proposed a supervised and unsupervised approach to detect learning style automatically [30]. They used a *k*-means clustering algorithm to group features into 16 learning style combinations based on FSLSM and then used the naïve Bayes classifier to predict the learning style, with an accuracy of 89%.

Recent research by Muhammad et al. proposed GRL-LS, an unsupervised approach to learning style detection [31]. They used a *k*-means clustering algorithm to identify learning style groups for learners based on FSLSM and experimented with KDDCup datasets with an average accuracy and precision of 88.25% and 78.50%, respectively. Rasheed and Wahid proposed a supervised approach to detect learning style automatically [32]. They used multiple classification algorithms and compared their accuracy on a dataset of 200 students. The highest accuracy was achieved with Support Vector Machine (SVM) classifier at 75.55%, while the lowest was with a decision tree at 45.55%. Similarly, Aziz et al. proposed the AFCM model, an unsupervised approach to learning style detection [33]. They used a Fuzzy C Mean clustering algorithm on a dataset of 249 students. Also, Mehenaoui et al. [34] proposes an automatic approach to identify learners' learning styles based on patterns of learning behavior using the FSLSM in an online learning environment. Various machine learning techniques were used to detect learners' learning styles, and experiments with 73 students were used to validate the approach's effectiveness. Varying levels of accuracy were observed for the different dimensions of FSLSM. Notably, the SVM classifier displayed improved ability to predict learning styles for all dimensions of FSLSM, achieving an average accuracy of 88%. In a similar approach, Altamimi et al. [35] proposes a probabilistic approach for predicting preferred learning styles. By employing five machine learning algorithms (e.g., multi-layer perceptron, SVM, Decision Trees, Random Forest, and K-nearest neighbor), the work predicts the probability of learning styles in a sample of 72 students. The results indicate that regression algorithms are more representative in predicting learning style probabilities. Others employed a Convolutional Neural Network-based Levy Flight Distribution algorithm for learning style prediction in an e-learning environment [36]. The algorithm was evaluated using data from the CAROL platform. The proposed methodology was applied for predicting different learning types, including Active/Reflective, Sensing/Intuitive, Visual/Verbal, and Sequential/Global with an accuracy of 85.32%, 90.47%, 78.98%, and 90.37%, respectively.

However, previous research on learning style detection has several limitations, including varying accuracy across different learning styles, small dataset size, lack of generalization, and different experimental protocols and course characteristics [3, 18, 25, 27-29, 31-34, 36]. To address these limitations, our research focuses on obtaining behavior data of students while interacting with Moodle LMS and classifying behavior based on their learning styles according to FSLSM. To enhance accuracy, we utilize a semi-supervised learning self-training algorithm, which learns based on labeled and unlabeled data. This approach enables us to improve the generalization and performance of our model.

## IV. RESEARCH METHODOLOGY

In our research, we propose a novel approach that integrates weblog mining strategies and machine learning algorithms to create a model that considers existing student behaviors. This model is designed to identify different learning styles based on the data collected from student behavior. To accomplish this goal, we introduce a five-phase methodology that includes data collection, data pre-processing, supervised learning training using labeled data, automated labeling of unlabeled data, and performance evaluation of the classification process.

### A. Proposed Methodology

The proposed methodology involves self-training for semi-supervised learning to produce a reliable classification model with a limited amount of labeled data. The approach involves training the model on a combination of labeled and unlabeled data, where typically only a small amount of labeled data is available.

To overcome the costly labeling process in supervised learning and the limited application spectrum of unsupervised learning, the proposed method utilizes the combination of both labeled and unlabeled data. The unlabeled data is used to predict labels using the trained model, creating self-taught labels. The newly labeled data is then combined with the labeled data to

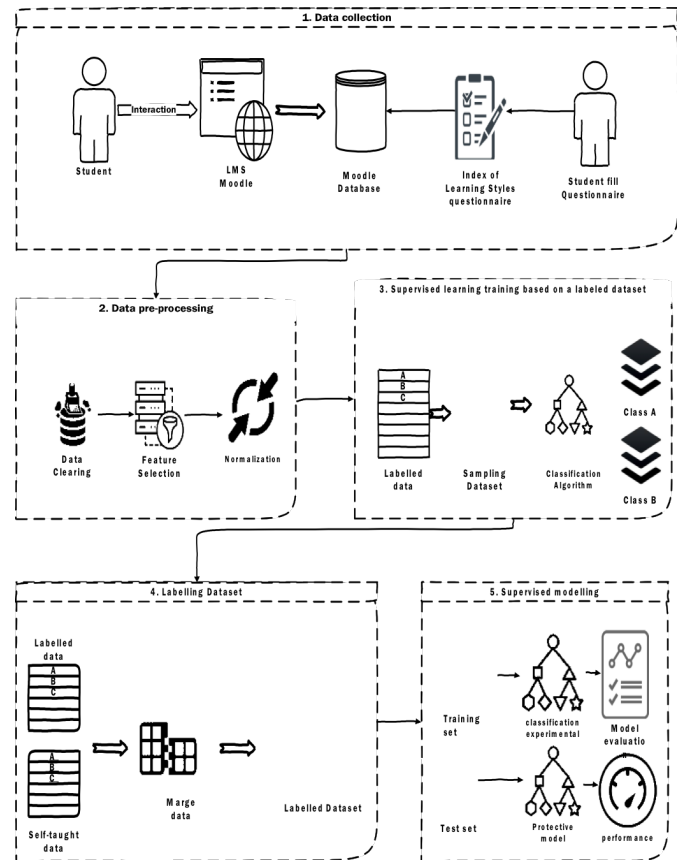

**Fig. 2:** Overview of the proposed method.

![Dataset before cleaning table]

**Fig. 3:** Dataset before data cleaning.

create a new dataset, which is used to train the model, see Fig. 2.

The proposed self-training for labelling unlabeled data algorithm, shown in Algorithm 1, takes weblog files of each learner as input and outputs the learning style of each learner in the four dimensions of the FSLSM model. The algorithm involves preprocessing the weblog files, creating a labeled and unlabeled dataset, sampling the labeled set, training the labeled dataset using a SVM classifier, classifying the unlabeled dataset using the trained model, and combining the newly labeled data with the labeled data to create a new dataset. Finally, the new data set is used to train the model. The proposed method offers a cost-effective approach to training machine learning models with limited labeled data, thereby increasing the potential for successful classification of data.

1) *Data Pre-processing*

Data pre-processing is a critical step that aims to enhance the quality of educational data before analysis. Although pre-processing in other domains may share similarities with educational data, there are certain characteristics that set it apart. Firstly, the vast amount of student information generated daily from multiple educational systems is a defining feature. Secondly, incomplete data is a common issue in education as most students do not finish all their activities and tasks, resulting in missing data. To address this, the pre-processing phase involves a series of steps that include data gathering, cleaning, reduction, filtering, and transformation.

The first step in pre-processing is data gathering, which involves collecting all available data from various educational systems generated at separate times and places. For this study,

**Fig. 4:** Dataset after data cleaning.

**Algorithm 1** Self-training for Labelling Unlabelled Data
1: **Input:** Weblog files of each learner
2: **Output:** Learning style of each learner in four dimensions of FSLSM
3: Access the weblog files of each learner.
4: Pre-process weblog files (Filter the activities and their corresponding actions of each learner from weblog files based on the four dimensions of FSLSM model)
5: Create dataset $D$
6: Split the dataset $D$ labeled set $L$ and the unlabeled set $U$
7: Sampling $L$
8: $TL = TRAIN(L, SVM)$ ▷ Train labeled dataset using SVM
9: $U' = Classify(TL, U)$ using $TL$ ▷ Labeling the unlabeled dataset based on Predict model
10: $D' = L + U'$ ▷ Combine the labeled data and the newly Self-taught-labeled data in a new dataset $D'$
11: $TRAIN(D', Supervisedmodel)$
12: **Output:** Train new dataset $D'$

**Fig. 5:** Data transformation example on data extracted from Moodle course management system.

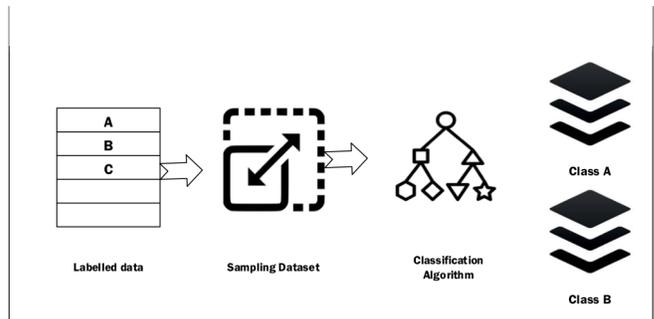

**Fig. 6:** Overview of supervised learning/training based on a labeled dataset.

data was collected from the Moodle database, and Table I provides a list of key features.

TABLE I
MOODLE DATABASE COMMON FEATURES

| Attribute | Description |
|---|---|
| **el_user** | all about user details |
| **el_log** | user's action |
| **el_assignment** | Data about every task |
| **el_assignment_submissions** | Data about tasks submitted |
| **el_forum** | Data about all forums |
| **el_forum_posts** | Data for all posts |
| **el_forum_discussions** | Data for all discussions |
| **el_message** | Stores all data about messages |
| **el_ message_reads** | Stores all data about the read messages |
| **el_quiz** | data about quizzes |
| **el_quiz_attempts** | Stores all data about quiz attempts |
| **el_quiz_grades** | Stores the final quiz grade |
| **el_youtube** | Logs every user's action when watching a video |
| **el_tag** | Information on all Learning object tags |

Data cleaning involves detecting erroneous or irrelevant data and discarding it [37]. In educational data, missing data is a common issue that often arises when students do not complete all the activities in the course. In some cases, students who have not completed all the activities can be removed from the data. The log file of each course contains fields such as 'Time,' 'User Full Name,' 'Affected User,' 'Event Context,' 'Event Name,' 'Description,' 'Origin,' and 'IP Address.' Some irrelevant fields such as 'IP Address' are removed. Fig. 3 provides a screenshot before data cleaning, and Fig. 4 provides a screenshot after data cleaning.

Data reduction involves selecting an appropriate subset of attributes and ignoring irrelevant and redundant ones [4]. For this work, useful data was chosen to detect learning styles. While the number of visits to video materials is essential, the identity of the individual responsible for adding the video remains unnecessary.

Data filtering involves selecting certain subsets of events, students, or courses, based on useful data to detect learning styles. The required data for this study was determined by mapping between learning style and learner's behaviors, using the Felder-Silverman learning style model. The preferences of the Visual Learner archetype are associated with learning materials presented in video format, as captured in the Moodle LMS database by tracking the frequency of visits to video materials.

Data transformation involves aggregating the total number of interactions for each student to each learning object, such as the number of visits and the number of posts. Fig. 5 provides a screenshot after data pre-processing.

By thoroughly undertaking these steps of data gathering, cleaning, reduction, filtering, and transformation, the data can be transformed into valuable insights for researchers and practitioners in education. The unique characteristics of educational data make it essential to undertake these steps thoroughly, so that the data is reliable and can be used for subsequent analysis.

### B. Supervised Learning/Training Based on a Labeled Dataset

This part describes the process of building a classifier using a labeled dataset for supervised learning. We begin by preprocessing the dataset, followed by inputting it into a classification algorithm. Fig. 6 provides an overview of the supervised learning/training process based on a labeled dataset.

#### 1) Preprocessing the Dataset

The first step in building a classifier is to preprocess the dataset. In this phase, we use labeled data to predict unlabeled data. Therefore, we follow a two-step process. First, we split the dataset into labeled and unlabeled data. Second, we sampled the labeling dataset. However, when collecting data about learning styles, we often come across a term called imbalanced class distribution. This phenomenon occurs when one of the classes is much higher or lower than any other class. To even up the classes, there are two main methods that one can use for sampling the dataset. The first method is called random over-sampling, where we add copies of instances from the underrepresented class. The second method is called random under-sampling, where we delete instances from the overrepresented class. In our case, we used the random under-sampling method for sampling the labeled dataset.

#### 2) Classification Model

Once the dataset is preprocessed, we use a classification algorithm to train the model to predict labels on the unlabeled data, thus creating self-taught labels (i.e., newly labeled data). The number of unlabeled samples is often larger than the number of labeled samples, which is a common challenge in supervised learning where we do not have enough labeled data. Therefore, adding cheap and abundant unlabeled data to labeled data is a useful technique to overcome this problem. We used SVM classifier to predict new labels. SVM is a popular algorithm in machine learning because it is accurate, robust, and computationally efficient. Furthermore, many works compare classifiers, and SVM is often among the best.

The results obtained from the classification of learning styles are outlined as learning styles for each learner. To determine a learning style, we must determine four dimensions: the processing dimensions, which are "Active" and "Reflective"; the input dimensions, which are "Visual" and "Verbal"; the understanding dimensions, which are "Sequential" and "Global"; and the perception dimensions, which are "Sensing" and "Intuitive". By considering these four dimensions, we can determine the learning style of a learner, which can then be used to tailor the learning experience to their specific needs.

### C. Labeling Dataset

Labeling is a crucial stage in the supervised learning process that helps us identify the learning style of each learner. It is the process of assigning labels that reflect the learning style for each dimension, allowing us to categorize learners based on their learning preferences. This categorization enables us to personalize the learning experience for each learner, which can lead to better learning outcomes.

After the supervised learning algorithm has been trained on a labeled dataset, our classification algorithm can categorize each pair of learners and learning styles using labels that reflect each dimension. These labels can include categories such as "active" or "reflective" for processing dimensions, "visual" or "verbal" for input dimensions, "sequential" or "global" for understanding dimensions, and "sensing" or "intuitive" for perception dimensions. The classification algorithm assigns a label to each pair, creating a labeled dataset that can be used to personalize the learning experience. Fig. 7 depicts the process of combining the labeled data with the newly self-taught-labeled data to create a new dataset used to train the model. By adding the newly labeled data to the original labeled dataset, we increase the size of the dataset, which can improve the accuracy of the classification algorithm.

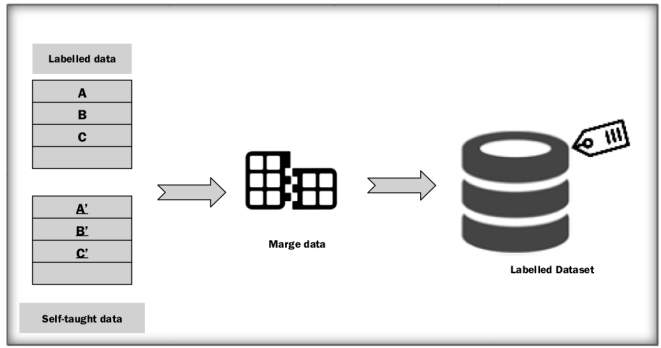

**Fig. 7:** Combine labeled data with newly self-taught-labeled data in a joint dataset.

### D. Supervised Modelling

Our methodology for personalized learning involves several stages, beginning with data collection and preprocessing, followed by supervised learning training based on a labeled dataset, labeling, and supervised modelling. In this part, we focus on the final stage of our methodology, which involves constructing a labeled dataset that can be used to feed the classifier algorithms.

As previously mentioned, labeling is a critical step in our methodology that allows us to assign labels to each learner based on their learning style dimensions. This process enables us to create a labeled dataset that can be used to train our classifier algorithms, which in turn can help personalize the learning experience for each learner. Supervised modelling involves using this labeled dataset to build a model that can predict the learning style of a new learner. This model takes as input the four dimensions of the learner's learning style, namely processing, input, understanding, and perception. The model then outputs a prediction of the learner's learning style for each of these dimensions.

One of the main advantages of supervised modelling is that it can be used to personalize the learning experience for each learner. By predicting the learning style of a new learner, we can tailor the educational content to match their preferred learning style, leading to improved learning outcomes. Moreover, the model can be updated periodically based on new data, ensuring that the personalized learning experience remains up-to-date and relevant.

## V. EXPERIMENTS AND RESULTS

In this section, we will outline the experiments we conducted and the results we obtained. To implement our proposed approach, we utilized Weka 3.8 and My SQL 5.6. Our experiments were conducted on a Microsoft Windows 10 Pro 64-bit operating system, equipped with a 2.70 GHz Intel Core i7 processor and 16 GB of memory.

### A. Data Collection

The current study utilized data collected from two courses offered on the University of Jordan's E-learning Portal, namely "Computer Skills for Humanities Students" and "Computer Skills for Medical Students" in the second semester of 2020. Throughout these courses' sixteen-week duration, students engaged in activities such as reading materials, video lectures, assignments, quizzes, and more. The log activity of each student was recorded and anonymized to preserve their privacy. To acquire the necessary data, we aggregated information from various sources such as the LMS and the student information system and consolidated it into a single database. Table II provides a summary of the data collection process employed in the experiments. The complete two course dataset can be downloaded from IEEE DataPort[1].

For this research, we opted to utilize the FSLSM. At the onset of the course, students completed the ILS, a 44-item questionnaire that gauge learners' personal preferences for each dimension by assigning values ranging from -11 to +11 per dimension based on the responses to eleven questions per dimension. The ILS questionnaire lets us capture students' preferences regarding the FSLSM dimensions. Table III displays the statistics derived from the completed questionnaires, including the number of students who participated in the survey.

TABLE II
DATASETS DETAILS

| Course | Learners | Events | Launch date | End date |
|---|---|---|---|---|
| Computer skills for humanities (CSFH) | 1749 | 1139810 | 1/21/2020 | 5/20/2020 |
| Computer skills for medical students (CSFM) | 564 | 484410 | 1/21/2020 | 5/20/2020 |

TABLE III
NUMBER OF STUDENT PARTICIPANTS IN THE QUESTIONNAIRE

| Course | Participants |
|---|---|
| Computer skills for humanities (CSFH) | 480 |
| Computer skills for medical students (CSFM) | 256 |

### B. Experimental Setup and Results

This experiment assessed the effectiveness of the SVM classifier in labeling unlabeled data using labeled data. The experiment was carried out on all dimensions of the learning style, including perception, processing, input, and understanding. In accordance with the guidelines, each run was performed on all dimensions of the dataset as described in Table II (containing labeled and unlabeled data). An SVM classifier was employed for each experiment, and the accuracy of the proposed method was measured using a confusion matrix. To assess the internal performance of the model, training was conducted using $k$-fold cross-validation with $k = 10$. Although the applied SVM classifier is known for its proven performance in various classification and ability to handle complex datasets[38-41], we have empirically experimented with other types of well-known classifiers and presented the results in Table IV to justify the choice of the SVM classifier for data labelling.

---
[1] https://dx.doi.org/10.21227/7tc4-5841.

TABLE IV
COMPARISON OF VARIOUS CLASSIFIERS

| Dataset | Dimension | SVM | RF | NB |
|---|---|---|---|---|
| Computer skills for humanities (CSFH) | Input | 0.84 | 0.87 | 0.82 |
| | Perception | 0.91 | 0.89 | 0.57 |
| | Processing | 0.95 | 0.99 | 0.93 |
| | Understanding | 0.71 | 0.71 | 0.70 |
| | Median | 0.88 | 0.88 | 0.76 |

The results of the labeling dataset experiment for the course of Computer Skills for Humanities are presented in Table V, which summarizes the correctly and incorrectly classified instances, including specificity, precision, recall, and Area Under the Receiver Operating Characteristic Curve (AUC-ROC). Based on the results of different dimensions, the processing dimension achieved the highest accuracy with 94.79%, while the understanding dimension obtained the lowest accuracy with 70.73%. The degraded performance in the understanding dimension may be attributed to the limited number of learning objects to detect dimensions.

TABLE V
RESULTS OF LABELING COMPUTER SKILLS FOR HUMANITIES

| Input Dimension | |
|---|---|
| Evaluation Metrics | Value |
| Correctly Classified Instances | 83.61% |
| Incorrectly Classified Instances | 16.39% |
| Specificity | 0.8 |
| Precision | 0.84 |
| Recall | 0.84 |
| AUC – ROC | 0.69 |
| **Perception Dimension** | |
| Evaluation Metrics | Value |
| Correctly Classified Instances | 88.76% |
| Incorrectly Classified Instances | 11.24% |
| Specificity | 0.91 |
| Precision | 0.91 |
| Recall | 0.89 |
| AUC – ROC | 0.71 |
| **Processing Dimension** | |
| Evaluation Metrics | Value |
| Correctly Classified Instances | 94.79% |
| Incorrectly Classified Instances | 5.20% |
| Specificity | 0.99 |
| Precision | 0.95 |
| Recall | 0.95 |
| AUC – ROC | 0.74 |
| **Understanding Dimension** | |
| Evaluation Metrics | Value |
| Correctly Classified Instances | 70.73% |
| Incorrectly Classified Instances | 29.27% |
| Specificity | 0.75 |
| Precision | 0.71 |
| Recall | 0.71 |
| AUC – ROC | 0.59 |

Similarly, the results of the labeling dataset experiment for Computer Skills for Medical Students are presented in Table VI, which summarizes the correctly and incorrectly classified instances, including specificity, precision, recall, and AUC-ROC. Based on the results of different dimensions, the input dimension obtained the highest accuracy with 84.41%, while the understanding dimension obtained the lowest accuracy with 47.98%.

Table VI
RESULTS OF LABELING COMPUTER SKILLS FOR MEDICAL STUDENTS

| Input Dimension | |
|---|---|
| Evaluation Metrics | Value |
| Correctly Classified Instances | 84.41% |
| Incorrectly Classified Instances | 15.59% |
| Specificity | 0.93 |
| Precision | 0.85 |
| Recall | 0.83 |
| AUC – ROC | 0.69 |
| **Perception Dimension** | |
| Evaluation Metrics | Value |
| Correctly Classified Instances | 52.36% |
| Incorrectly Classified Instances | 47.64% |

| Evaluation Metrics | Value |
|---|---|
| Specificity | 0.52 |
| Precision | 0.52 |
| Recall | 0.52 |
| AUC – ROC | 0.51 |
| **Processing Dimension** | |
| Evaluation Metrics | Value |
| Correctly Classified Instances | 48.94% |
| Incorrectly Classified Instances | 51.06% |
| Specificity | 0.50 |
| Precision | 0.48 |
| Recall | 0.49 |
| AUC – ROC | 0.49 |
| **Understanding Dimension** | |
| Evaluation Metrics | Value |
| Correctly Classified Instances | 47.98% |
| Incorrectly Classified Instances | 52.02% |
| Specificity | 0.48 |
| Precision | 0.48 |
| Recall | 0.48 |
| AUC – ROC | 0.49 |

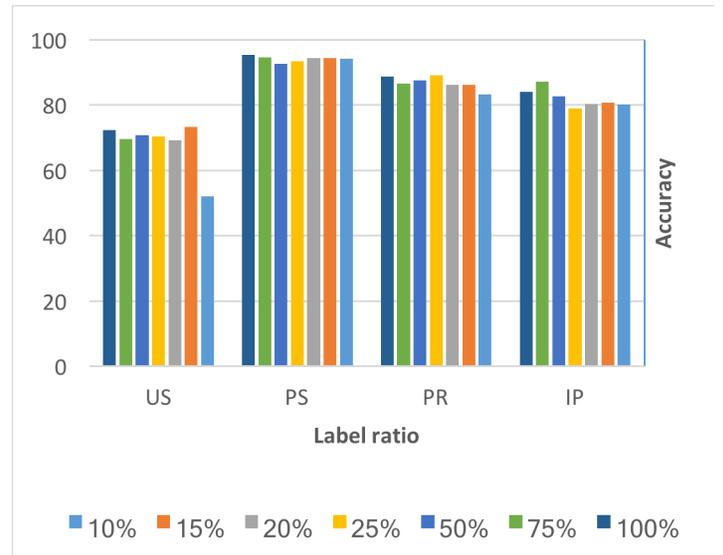

**Fig. 8:** Different ratios of labeled data used for training the SVM classifier on the CSFH dataset showing understanding (US), processing (PS), perception (PR) and input (IP) dimensions, respectively.

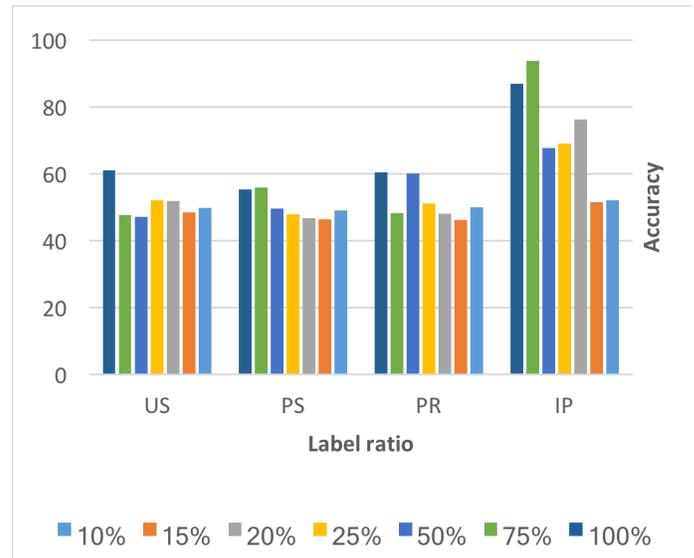

**Fig. 9:** Different ratios of labeled data used for training the SVM classifier on the CSFM dataset showing understanding (US), processing (PS), perception (PR) and input (IP) dimensions, respectively.

It is worth noting that the results on the two datasets are dissimilar, and there is a variance of the results on the same dataset. For instance, in the first dataset ("Computer Skills for Humanities Students"), the processing dimension showed the highest accuracy of 94.8%, while the understanding dimension showed the lowest accuracy of 70.7%. These observations suggest that the SVM classifier's effectiveness in labeling unlabeled data using labeled data may vary depending on the dataset's characteristics and dimensions.

Labeled ratio in semi-supervised machine learning defines the proportion of labeled data to the total dataset – including both labeled and unlabeled instances – used for model training. It is a pivotal parameter that influences how the model learns from labeled data and capitalizes on the information from unlabeled data. A higher labeled ratio enhances labeled data utilization and may mitigate overfitting, while a lower ratio shows the potential of a larger pool of unlabeled data for improved generalization. In our work, we investigated five distinct labeled ratios: 0.1, 0.2, 0.5, 0.75, and 1. Fig. 8 and Fig. 9 show results with different labeled ratios. The labeled ratios classification accuracies are consistent across all learning styles dimensions for the CSFH dataset (c.f. Fig. 8). Similarly applies to the CSFM dataset in Fig. 9, however some heterogeneity is evident in the IP dimension.

On the other hand, the experiment with the proposed method aims to assess the classifier's effectiveness in identifying learning styles. The experiment adhered to a set of guidelines that included testing all dimensions listed in Table II, using various classifiers such as RF, SVM, C4.5, and Neural Network (NN), measuring accuracy using a confusion matrix, and employing $k$-fold cross-validation with $k = 10$ for model training. To evaluate the performance, all dimensions, including perception, processing, input, and understanding. For the Computer Skills for Humanities dataset, we selected SVM, Neural Network, Random Forest, and J48 Decision Tree as the classification models, and the results of their performance are presented in Table VII. The results are summarized based on correctly and incorrectly classified instances, including specificity, precision, recall, and AUC – ROC.

TABLE VII
RESULTS OF SELF-TRAINING OF SEMI-SUPERVISED LEARNING
METHODS FOR COMPUTER SKILLS FOR HUMANITIES DATASET

| Input Dimension | | | | |
|---|---|---|---|---|
| Evaluation Metrics | SVM | NN | RF | J4.8 |
| Correctly Classified Instances | 96.86% | 96.92% | 97.54% | 97.54% |
| Incorrectly Classified Instances | 3.14% | 3.08% | 2.46% | 2.46% |
| Specificity | 0.76 | 0.76 | 0.79 | 0.79 |
| Precision | 0.97 | 0.97 | 0.98 | 0.98 |
| Recall | 0.97 | 0.97 | 0.98 | 0.98 |
| AUC – ROC | 0.94 | 0.99 | 0.99 | 0.97 |
| **Perception Dimension** | | | | |
| Evaluation Metrics | SVM | NN | RF | J4.8 |
| Correctly Classified Instances | 98.20% | 98.38% | 98.42% | 98.28% |
| Incorrectly Classified Instances | 1.79% | 1.64% | 1.58% | 1.72% |
| Specificity | 0.77 | 0.81 | 0.83 | 0.83 |
| Precision | 0.99 | 0.99 | 0.99 | 0.98 |
| Recall | 0.98 | 0.98 | 0.98 | 0.98 |
| AUC – ROC | 0.99 | 0.99 | 0.99 | 0.95 |
| **Processing Dimension** | | | | |
| Evaluation Metrics | SVM | NN | RF | J4.8 |
| Correctly Classified Instances | 97.48% | 97.48% | 99.06% | 98.89% |
| Incorrectly Classified Instances | 2.52% | 2.52% | 0.94% | 1.11% |
| Specificity | 0.90 | 0.85 | 0.96 | 0.95 |
| Precision | 0.98 | 0.98 | 0.99 | 0.99 |
| Recall | 0.98 | 0.98 | 0.99 | 0.99 |
| AUC – ROC | 0.94 | 0.99 | 0.99 | 0.95 |
| **Understanding Dimension** | | | | |
| Evaluation Metrics | SVM | NN | RF | J4.8 |
| Correctly Classified Instances | 93.75% | 93.68% | 92.97% | 93.49% |
| Incorrectly Classified Instances | 6.25% | 6.32% | 7.03% | 6.51% |
| Specificity | 0.64 | 0.62 | 0.62 | 0.65 |
| Precision | 0.94 | 0.95 | 0.94 | 0.93 |
| Recall | 0.94 | 0.94 | 0.93 | 0.94 |
| AUC – ROC | 0.83 | 0.97 | 0.97 | 0.91 |

Similarly, for the Computer Skills for Medical Students dataset, we employed the same classifiers to test classification performance across all four dimensions of the FSLSM. The results of the self-training of semi-supervised learning method experiment are presented in Table VIII, and they are also summarized based on correctly and incorrectly classified instances, including specificity, precision, recall, and AUC – ROC. Table IX displays the statistical significance of comparing supervised and semi-supervised (self-taught) learning. We used a paired T-test, as recommended by previous studies [42], to assess performance metrics (e.g. accuracy, precision, and recall) on the same dataset for both approaches.

TABLE VIII
RESULTS OF SELF-TRAINING FOR SEMI-SUPERVISED LEARNING
METHODS FOR COMPUTER SKILLS FOR MEDICAL STUDENTS

| Input Dimension | | | | |
|---|---|---|---|---|
| Evaluation Metrics | SVM | NN | RF | J4.8 |
| Correctly Classified Instances | 94.76% | 95.03% | 96.60% | 96.60% |
| Incorrectly Classified Instances | 5.24% | 4.97% | 3.40% | 3.40% |
| Specificity | 0.96 | 0.98 | 0.99 | 0.99 |
| Precision | 0.95 | 0.95 | 0.97 | 0.97 |
| Recall | 0.95 | 0.95 | 0.97 | 0.97 |
| AUC – ROC | 0.79 | 0.99 | 0.98 | 0.99 |
| **Perception Dimension** | | | | |
| Evaluation Metrics | SVM | NN | RF | J4.8 |
| Correctly Classified Instances | 74.73% | 73.82% | 76.00% | 72.73% |
| Incorrectly Classified Instances | 25.27% | 26.18% | 24.00% | 27.27% |
| Specificity | 0.75 | 0.79 | 0.80 | 0.76 |
| Precision | 0.77 | 0.71 | 0.74 | 0.68 |
| Recall | 0.75 | 0.74 | 0.76 | 0.73 |
| AUC – ROC | 0.52 | 0.73 | 0.79 | 0.70 |
| **Processing Dimension** | | | | |
| Evaluation Metrics | SVM | NN | RF | J4.8 |
| Correctly Classified Instances | 78.06% | 74.93% | 75.37% | 76.57% |
| Incorrectly Classified Instances | 21.94% | 25.07% | 24.63% | 23.43% |
| Specificity | 0.77 | 0.69 | 0.73 | 0.74 |
| Precision | 0.78 | 0.75 | 0.75 | 0.76 |
| Recall | 0.78 | 0.75 | 0.75 | 0.77 |
| AUC – ROC | 0.75 | 0.75 | 0.73 | 0.73 |
| **Understanding Dimension** | | | | |
| Evaluation Metrics | SVM | NN | RF | J4.8 |
| Correctly Classified Instances | 72.71% | 73.04% | 72.71% | 70.59% |
| Incorrectly Classified Instances | 27.29% | 26.96% | 27.29% | 29.41% |
| Specificity | 0.65 | 0.68 | 0.68 | 0.64 |
| Precision | 0.77 | 0.74 | 0.77 | 0.73 |
| Recall | 0.72 | 0.73 | 0.72 | 0.7 |
| AUC – ROC | 0.74 | 0.80 | 0.79 | 0.72 |

The results indicate that the proposed method performs effectively in identifying learning styles across different dimensions. The use of various classifiers yielded varying results, indicating that different classifiers may perform better or worse depending on the dataset and the specific dimensions being evaluated. Nonetheless, the proposed method can serve as a promising tool for labeling unlabeled data, especially when sufficient labeled data is not available.

In addition to the applied self-taught method, our experimental setup also includes comparison with other semi-supervised methods such as tri-training. Fig. 10 and Fig. 11 show a comparison of results with different labeled ratios. This comparative analysis investigates the varying performance of the semi-supervised approaches across the used datasets. The proposed self-training approach shows much better performance on the perception dimension when applied to the CSFH dataset (see Fig. 10), and appears better on the perception dimension, slightly less effective on the processing dimension, and consistent to the tri-training approach for the CSFM dataset (see Fig. 11).

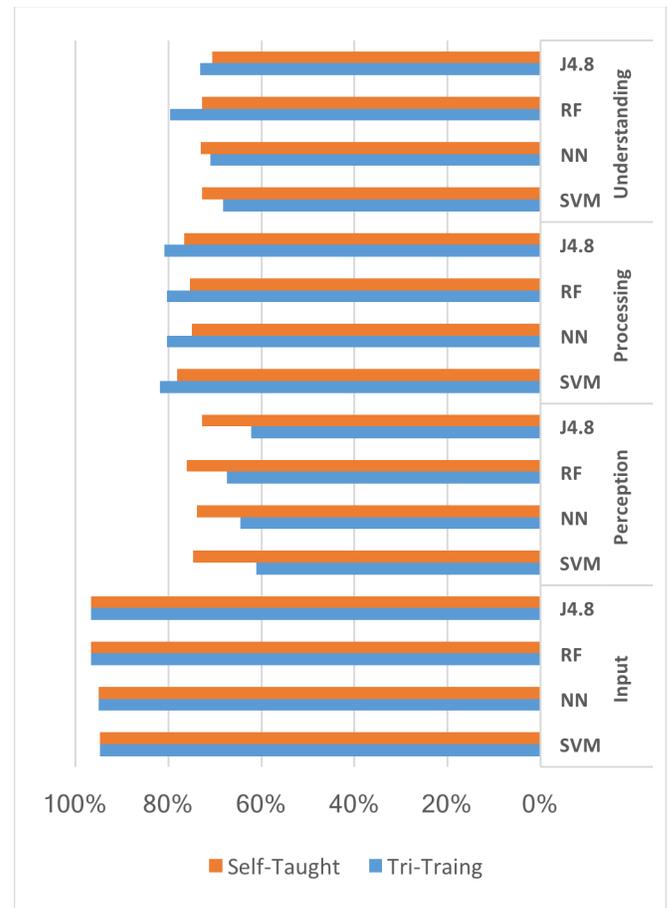

**Fig. 11:** Performance evaluation of Tri-training vs. Self-Taught on CSFM dataset.

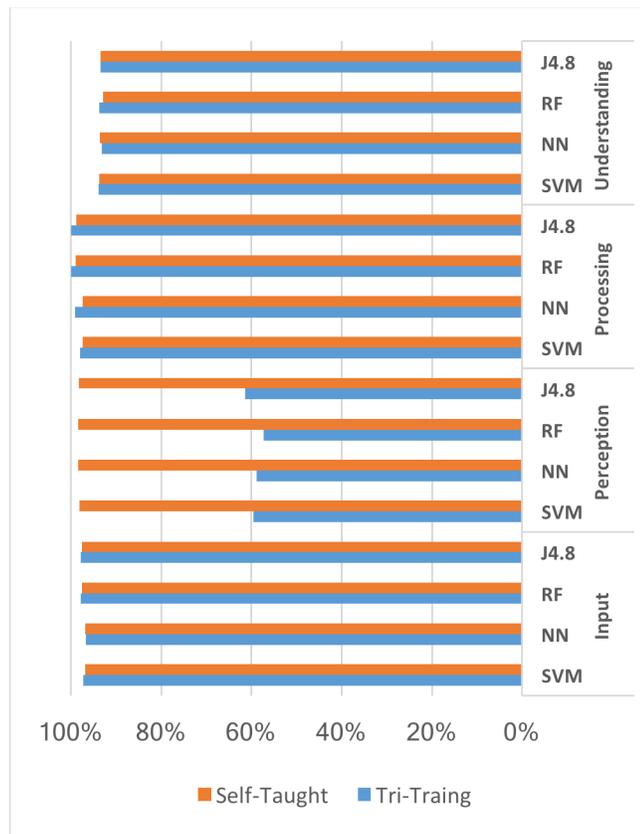

**Fig. 10:** Performance evaluation of Tri-training vs. Self-Taught on CSFH dataset.

TABLE IX
SUPERVISED LEARNING VS. SEMI-SUPERVISED LEARNING (SELF-TAUGHT).

| | **T-test paired** | | |
|---|---|---|---|
| **Datasets** | **Matric** | **SL vs SSL** | |
| | | **T-value** | **P-value** |
| CSHS | Accuracy | -3.20 | 0.041 |
| CSHS | Precision | -2.87 | 0.033 |
| CSHS | Recall | -3.75 | 0.014 |
| CSMS | Accuracy | -5.38 | 0.011 |
| CSMS | Precision | -5.07 | 0.014 |
| CSMS | Recall | -6.14 | 0.008 |

## VI. ANALYSIS AND INTERPRETATION

We applied Algorithm 1 to predict learning styles, and observed dissimilar results on two datasets, namely "Computer skills for humanities Students" and "Computer skills for medical students." Additionally, we noted variance of the results on the same dataset, where the processing dimension showed the highest accuracy (94.8%), and the understanding dimension showed the lowest accuracy (70.7%) for the first dataset. We attribute these results to differences in course content design and dataset size. We then investigated the results for the learning style dimension using self-training for semi-supervised learning with four different classifiers. Cross-validation (10-fold) was used to train and test the models, where the dataset was randomly split into '10' groups, and one group was used as the test set while the rest were used as the training set. We found that all machine learning classifiers provided satisfactory results for all dimensions; however, random forest (RF) and C4.5 (J48) showed the highest accuracy value, making them the most appropriate for our purposes. These classifiers offered the best prediction performance, indicating that they are most suitable for predicting the degree of preference of each learning style dimension shown by learners during their interactions with the Moodle platform.

Our datasets contained both labeled and unlabeled data, with the labeled data being small and highly imbalanced. To increase the quality of our classifiers, we labeled the data at some point. Our results in Tables VII and VIII support our finding that our approach is accurate and suitable for predicting learners' learning styles from supervised learning when the labeled data is limited.

The proposed model can be used in several ways, such as making learners aware of their learning styles and allowing them to identify their strengths and weaknesses. Also, the model can help teachers and instructors obtain insight into their learners' preferences and customize instructions accordingly. Furthermore, since the proposed model responds quickly to changes in the learner's learning style, it offers an important approach for enhancing adaptivity in e-learning platforms by enabling these systems to provide tailored instructions and activities. However, a question arises about the accuracy of the self-trained semi-supervised model. The performance of the classification problem is evaluated according to its accuracy, which is essential to know how close the predicted values are to the ground truth. One problem with self-training is that a supervised learning algorithm's mistakes reinforce themselves, hence, accuracy does not reflect an accurate measure concerning self-training. Therefore, we used the equation Accuracy = (labeled value + correct predictions)/total number to calculate the self-taught accuracy (Accuracy_ST) for the dataset.

We compared supervised learning and our semi-supervised learning approach and found that the latter is indeed effective over all dimensions for predicting learners' learning styles. Fig. 12 shows a comparison between the two approaches. Furthermore, when comparing our results to those from related work, our method presented the best result, with an accuracy of 88.3% against 89.9% achieved by the work of Maaliw [27].

However, the work of Maaliw considered only 108 students, while we considered 1705 students.

Our results presented in this research outperform previous works that explored the automatic detecting of Learning Styles and using Moodle LMS, as shown in Table X. Nevertheless, a comparative analysis of the precision achieved by other studies needs more investigation, as each work was performed using different experimental protocols, and course characteristics, rendering comparing different approaches across various datasets to be challenging. However, this work employs a consistent experimental process, and standardized methods were followed for extracting the course details. By investigating common ways of identifying prevalent features, the aim was to establish an unbiased foundation for evaluating diverse approaches used in relevant studies. Furthermore, the paper introduces a substantial dataset that has been extracted and preprocessed from a Learning Management System utilized within an academic institution (accessible at: https://dx.doi.org/10.21227/7tc4-5841). This dataset holds the potential to assist researchers in dealing with the complexities of discerning students' learning styles and can serve as a benchmark for conducting comparative analyses.

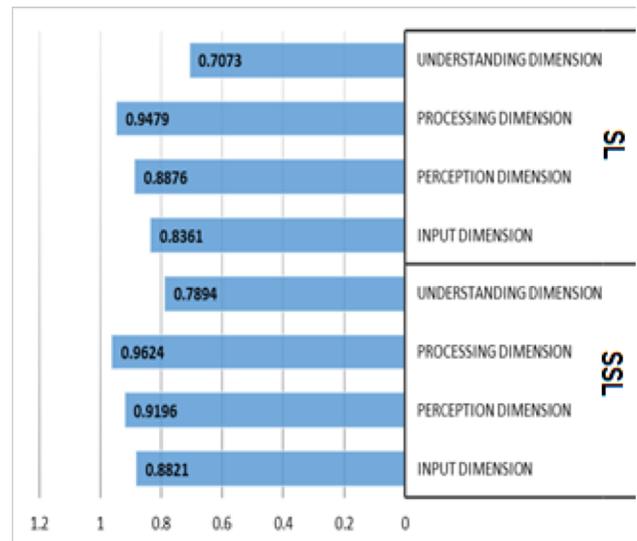

**Fig. 12:** Comparison between supervised learning (SL) and self-training learning (SSL).

Table X
BENCHMARKING WITH RELATED WORK

| Contribution | Accuracy |
|---|---|
| El Fazazi et al. [25] | 79.1% |
| Abdullah et al. [18] | 76.0% |
| Karagiannis and Satratzemi [3] | 72.8% |
| Maaliw [27] | 89.9% |
| Liyanage, Gunawardena, & Hirakawa [28] | 81.8% |
| García et al. [26] | 62.5% |
| **Our self-training approach** | 88.3% |

## VII. CONCLUSION

Personalized learning has shown the potential to enhance learning outcomes by tailoring educational experiences to individual strengths, needs, skills, and interests. In our work, we have proposed a novel self-training technique for identifying distinct learning styles. This technique was developed through the analysis of data from two courses offered at the University of Jordan's e-learning platform. Through a semi-supervised approach, we successfully constructed a robust classification model that effectively pinpointed the most suitable learning style for each student. Our proposed approach involves leveraging the grouping feature and lesson module within the Moodle platform to provide adaptive course content based on the identified learning styles. This approach can significantly enhance the effectiveness of LMS systems.

To further validate and refine our approach, there is a need for future research. This could involve applying our proposed adaptive system to diverse scenarios and testing its efficacy in real-time settings. Additionally, an experimental implementation of the adaptive system could be conducted to explore the correlation between different learning styles and students' academic performance. Such investigations could offer valuable insights into the practical implications and benefits of tailoring education to individual learning preferences.

ACKNOWLEDGMENT

The authors would like to thank Prof. Thair Hamtini for insightful comments, and Lama Rajab Abdel-Majeed and Aman Rahahleh for their support in collecting the dataset.

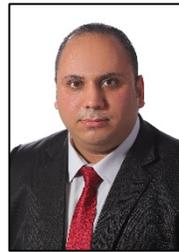
**Hany Y. Ayyoub** Hani Y. Ayyoub was born in Huson village, Irbid, Jordan in 1990. He received the B.S. degrees in Computer Science from the Al-Balqa' Applied University/HUC, Irbid, Jordan in 2012 and the M.S. degrees in Information Systems from University of Jordan, Amman, Jordan, in 2020.

Since 2014, he was an E-learning Administrator for the University of Jordan Campus until now. His research interests include eLearning, blended learning, flipped learning, project learning, collaborative learning, Mobile learning, adaptive learning, personalized learning, educational data mining, and Learning Analytics. He frequently identifies and assesses relevant learning technologies to extend, and support teaching, learning, plan, and manage the integration of such technologies into teaching practice.

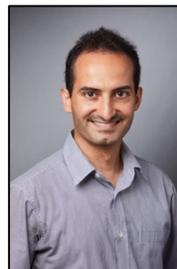
**Omar S. Al-Kadi** is a Professor in Computational Imaging and Machine Intelligence at the King Abdullah II School for Information Technology at the University of Jordan. He has a PhD from the University of Sussex in the UK and has held research positions at the University of Oxford, the University of Surrey, and as a visiting professor at the Swiss Federal Institute of Technology and Yale University. He has made several contributions to the field of pattern recognition and decision support systems for real-world problems, particularly in Artificial Intelligence in medicine. He has also applied nature-inspired algorithms to develop more intelligent algorithms based on real-world examples of social behavior. Dr. Al-Kadi is a senior member of the Jordan Engineers Association (JEA) and the Institute of Electrical and Electronic Engineers (IEEE).